\documentclass[12pt]{article}
\usepackage{amssymb}
\usepackage{subfigure}
\usepackage{graphicx}
\usepackage{float}

\usepackage{amsmath}
\usepackage{color}

\newcommand{\bb}{\begin{equation}}
\newcommand{\ee}{\end{equation}}
\newcommand{\bega}{\begin{eqnarray}}
\newcommand{\ega}{\end{eqnarray}}
\newcommand{\begae}{\begin{eqnarray*}}
\newcommand{\egae}{\end{eqnarray*}}

\newcommand{\h}{\hspace*{4ex}}

\newcommand{\cent}{\centerline}
\newcommand{\vs}{\vspace*}

\begin{document}

\baselineskip 0.5cm

\begin{center}

{\large {\bf Experimental optical trapping of micro--particles with higher--order Frozen Waves} }

\end{center}

\vs{0.2 cm}

\cent{Rafael A. B. Suarez$^{\: 1}$, Antonio A. R. Neves$^{\: 2}$, and Marcos R. R. Gesualdi$^{\: 2}$}

\vs{0.2 cm}

\centerline{{\em $^{\: 1}$ Facultad de Ciencias Básicas, Universidad Santiago de Cali, Campus Pampalinda, Santiago de Cali,  Colombia.}}
\centerline{{\em $^{\: 2}$ Universidade Federal do ABC, Santo Andr\'e, SP, Brazil.}}

\vs{0.5 cm}

{\bf Abstract  \ --} \ In this work, we optically trapping micro--particles with higher--order Frozen Wave using holographic optical tweezers. Frozen Waves are diffraction--resistant optical beams, obtained by superposing co--propagating Bessel beams with the same frequency and order, obtaining  efficient modeling of its shape. Based on this, we developed a holographic optical tweezers system for the generation of Frozen Waves and with this, it was possible to create traps in a stable way for the trapping and guiding of the micro--particles in the transverse plane. The experimental results show that it is possible to obtain excellent stability condition for optical trapping using higher--order Frozen Waves. These results indicate that the Frozen Waves is promising for optical trapping and guiding of particles, which may be useful in various application such as biological research, atomic physics and optical manipulations using structured light with orbital angular momentum. \\


\vs{0.5 cm}

\h {\em\bf 1. Introduction} 

Optical tweezers have become a powerful tool for application in different fields of research, mainly in the manipulation of biological system~\cite{Chiou2005,favre2019optical1,bustamante2021optical,cheppali2022forces}, colloidal systems~\cite{yada2004direct}, in nanotechnology for trapping of nano--structures~\cite{Marago2003,kotsifaki2022role}, in optical guiding and trapping of atoms~\cite{Kuga1997,vilas2022magneto} as well as the study of mechanical properties of polymers and biopolymers~\cite{sun2001stretching}. Initially, Ashkin \textit{et al}.~\cite{Askin1986} were able to capture in three--dimensional dielectric particles using a single beam well focused by a high numerical aperture lens. This allowed the measurement of picoNewton order forces on micro--sized particles with high precision. However, currently, most of its applications still involve the independent manipulation of multiple traps, which implies the generation of complex optical traps, thus requiring even more complex experimental systems~\cite{liesener2000multi,curtis2002dynamic}.
\\

Meanwhile, the study of non--diffracting waves or diffraction--resistant waves in optics has led to special beams that maintain their intensity pattern during propagation. Non--diffracting beams include Bessel beams, Airy beams, and others~\cite{Sivilo2007,Siviloglou2007,zhao2010propagation,Suarez2016}. Further, the superposition of these waves can also produce very special and interesting structured light beams~\cite{Vieira2012,suarez2019}. These special optical beams present great interest recently in optical tweezers, for trapping and guiding micro-- and nano–-particles~\cite{ambrosio2015optical, Suarez2020, suarezexperimental2020,suarez2021hotaab}.
\\

The introduction of holographic optical elements for the improved optical tweezers configuration has certainly become an advantage in the process of trapping, moving, and manipulating micro--particles with high precision~\cite{curtis2002dynamic,brzobohaty2013optical,Sun2008}. The use of computer--generated holograms and diffractive optical elements, such as spatial light modulators, allowed the simultaneous creation of several configurations of the optical traps, each with its unique characteristics~\cite{dufresne2001computer}. These trapping systems, called holographic optical tweezers, via wavefront control allows for easy three--dimensional positioning of traps, as well as the creation of special optical beams such as the Bessel, Laguerre--Gauss, Airy beams and a specific superposition of these beams, generating new special structured optical beams~\cite{Zhang2011,Zhang2013,suarezexperimental2020,suarez2021hotaab}.
\\

In this work, we present an experimental optical trap of micro--particles with higher--order FWs beams using holographic optical tweezers. Based on this, we investigate the optical force distribution acting on micro--particles using an optimized experimental holographic optical tweezers system. This system presents very interesting possibilities for static and dynamical applications to obtain greater stability for optical trapping micro--particles using higher--order FW and structured light with orbital angular momentum.
\\

\h {\em\bf 2. Frozen Waves} 

\textbf{Theory:} Frozen Waves (FWs) are resulting in an interesting method developed, capable to furnish non--diffracting beams whose longitudinal intensity shape can be freely chosen a priori ~\cite{Rached2004,Rached2005}. This approach is based on a suitable superposition of equal frequency and co--propagating Bessel beams. Besides a strong control on the longitudinal intensity pattern, this method also allows a certain control on the transverse shape of the resulting beam. A few years later, the experimental generation of the FW was achieved through computer--generated holograms reproduced by a spatial light modulator (SLM)~\cite{Vieira2012,Vieira2015}.
\\

Scalar Frozen Wave of $\nu$--order is given by~\cite{Rached2004}
\begin{equation}
\Psi\left(\rho,\phi,z\right)=\sum_{n=-N}^{N}A_{n}J_{\nu}\left(k_{\rho_{n}}\rho\right)e^{ik_{zn}z}e^{i\nu\phi}\,,
\label{FW_FW}
\end{equation}
where $k^2_{\rho_{n}}=k^2-k^2_{z_{n}}\,,$ $ k_{\rho n}=\sqrt{2k}\sqrt{\left(k-Q\right)-2 pi n/L}$ and $k_{zn}~=~Q~+~2\pi n/L$ are the transverse and longitudinal wave numbers, of the $2N+1$ Bessel beams in the superposition~\eqref{FW_FW}.
\\

In Eq.~\ref{FW_FW} the coefficients $A_{n}$ are defined from the desired longitudinal pattern, which is a function of $z$, these coefficients define the amplitude and the relative phase between the Bessel beams to have the chosen longitudinal profile, so
\begin{equation}
A_{n}=1/L \int_{0}^{L}F(z)e^{-i 2\pi nz/L}dz\,.
\label{An1}
\end{equation}
Once the choices for $k_{\rho n}$,  $k_{zn}$ and $A_{n}$ are made, we have that~\eqref{FW_FW} results in a beam with $|\Psi(\rho=0,z,t)| \approx |F(z)|^2$ where $|F(z)|^2$  is the desired longitudinal intensity pattern in the interval $0\leq z \leq L$ and with spot radius 

\begin{equation}
\Delta \rho \approx 2.4 \sqrt{k^2-Q^2}\,.
\label{Spot}
\end{equation}

The $Q$ parameter is a constant, where $0\leq Q \pm (2\pi/L)N \leq \omega/c$ for $-N\leq n \leq N$, which is important for the experimental conditions of FW generation ~\cite{Vieira2012,Vieira2015}.
\\

\textbf{Propagation of a higher--order FW through ABCD paraxial optical system:} The propagation of a FW through an ABCD paraxial optical system can be described by the generalized Huygens--Fresnel diffraction integral, which in cylindrical coordinates is given by~\cite{zhao2010propagation}
\begin{equation}
\begin{split}
\Psi & \left(\rho,\phi, z\right) = \dfrac{ik}{2\pi B}\int_{0}^{\infty}\int_{0}^{2\pi}  \rho_{0}d\rho_{0}d\phi_{0}\Psi\left(\rho_{0},\phi_{0} \right)\\
& \times \exp\left \lbrace \dfrac{-ik}{2B} \left[A\rho^{2}_{0} - 2\rho \rho_{0}\cos\left(\phi - \phi_{0}\right) + D \rho^{2} \right]  \right\rbrace \,,
\end{split}
\label{generalized_Huygens_Fresnel_diffraction_integral_FW}
\end{equation}
where $\rho_{0}$, $\phi_{0}$ and $\rho$, $\phi$ are the radial and azimuth angle coordinates in the input (SLM) and output planes, respectively. After some calculations, we get an analytical expression for the FW of $\nu$--order
\begin{equation}
\begin{split}
\Psi\left(\rho,z\right)&=\dfrac{(-1)^{\nu +1}}{A}\exp\left[-i\dfrac{B}{A}\left(k-Q \right)\right]\exp\left(i\nu \pi\right)\\
& \times \exp(ikz) \sum_{n=-N}^{N}A_{n}\exp\left(-i\dfrac{2\pi B}{AL}n \right) J_{\nu}\left(\dfrac{k_{\rho n} \rho}{A} \right).  
\end{split}
\label{generalized_Huygens_Fresnel_diffraction_integral_FW2}
\end{equation}

These results show that FW retains its properties when passing through an optical system, but characteristic parameters such as spot and longitudinal intensity profile can be modified.

\begin{figure}[htbp]
 \centering
  \includegraphics[width=\linewidth]{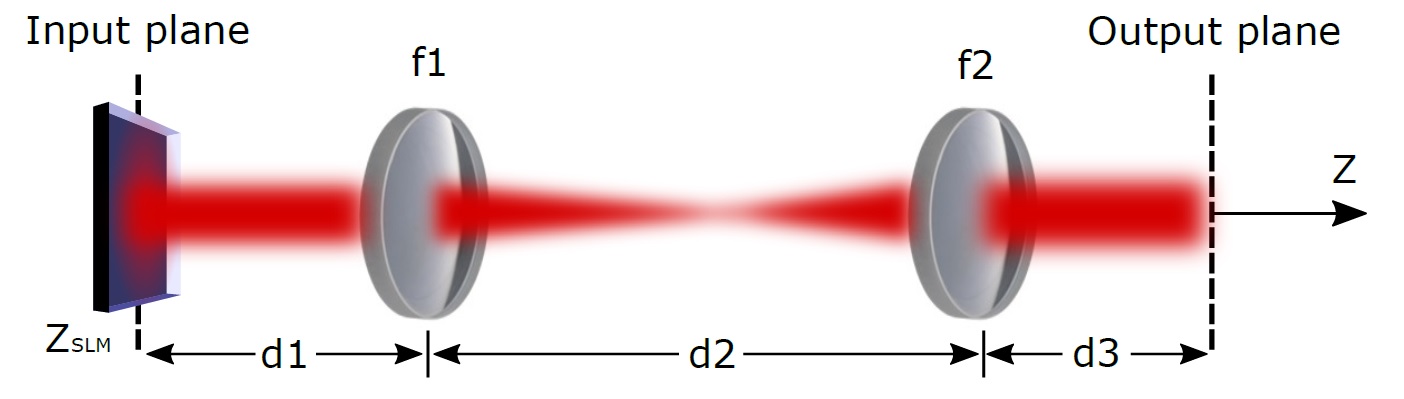}
 \caption{Lens configuration in a $4f$ Fourier filter system}
 \label{4f_system}
\end{figure}

The optical system shown in Fig~\ref{4f_system}, is a $4f$ imaging system, composed of two lenses of focal length $f_{1}$ and $f_{2}$, which results in a sequence of two Fourier transforms are placed directly after the input plane. The higher--order FW incidents in the input plane, and we observe the intensity distribution in the output plane. When the higher--order FW optically reconstructed by SLM pass through the  $4f $ systems, their propagation properties change according to the ~\eqref{generalized_Huygens_Fresnel_diffraction_integral_FW2}. For this system, the ABCD matrix is given by 
\begin{equation}
M=\begin{pmatrix}
A & B\\ 
C & D
\end{pmatrix} = \begin{pmatrix}
\dfrac{-f_{2}}{f_{1}} & \dfrac{-f_{1}}{f_{2}}z\\ 
0 & \dfrac{-f_{1}}{f_{2}}
\end{pmatrix} \,,
\label{ABCD-matrix}
\end{equation} 
%
%
%
%
%
%
where $d$ is the distance between the optical elements and $f$ is the focal length of the lenses. Where $d_{1}=f_{1}$, $d_{2}=f_1+f_2$, and $d_{3}=f_2+z$.
The \eqref{generalized_Huygens_Fresnel_diffraction_integral_FW2} becomes

\begin{equation}
\begin{split}
\Psi \left(\rho,z\right)&=(-1)^{\nu + 1}\left(\dfrac{f_{1}}{f_{2}}\right)\exp\left(i\nu \pi \right)\exp\left[-i\left(\dfrac{f_{1}}{f_{2}}\right)^2 \left(k-Q \right)z\right]\\ 
& \times \sum_{n=-N}^{N}A_{n}\exp\left(-i\dfrac{2\pi n}{(f_{2}/f_{1})^2 L}z \right) J_{\nu}\left(\dfrac{k_{\rho n} \rho}{(f_{2}/f_{1})} \right)\,,
\end{split}
\label{FW_OT_ABC}
\end{equation}
with

\begin{equation}
\begin{split}
A_{n}=\dfrac{1}{L(f_{2}/f_{1})^2} \int_{0}^{L(f_{2}/f_{1})^2}& F\left[\left(\dfrac{f_{1}}{f_{2}} \right)^2 Z \right]\\
& \times \exp\left[-i\dfrac{2\pi n}{(f_{2}/f_{1})^2 L}Z \right]dZ\,, 
\end{split}
\label{An}
\end{equation}
where $z=Z(f_{1}/f_{2})^2.$

From~\eqref{FW_OT_ABC} and \eqref{An}, the size of the beam spot changes with the focal distance ratio, $(f_{2}/f_{1})$, whereas the longitudinal pattern of intensity changes with the square of the focal length ratio $(f_{2}/f_{1})^2$~\cite{vittorino2019,rafaoptik2019}.
\\

\h {\em\bf 3. Holographic optical tweezers} 

\textbf{Computational holographic method for FW generation}
The computational holographic methods are now a well--established technique for the generation and characterization of special optical beams and structured light, particularly non--diffracting beams~\cite{Vieira2012,Vieira2015,Suarez2016,tarciofwoc14, tarcioprapb17, gesualdijoam06, Yepes2019, Gesualdi2008}. The Computer--Generated Hologram (CGH) of these optical beams are calculated and implemented in spatial light modulators and reconstructed optically in a holographic setup. These methods have experimentally generated optical beams of high quality and fidelity because the holographic technique is an extremely accurate tool in the reconstruction of amplitude and phase of optical waves~\cite{Vieira2012,Vieira2015,Suarez2016,suarez2019}. Particularly, in this work using the field that described the FW Eq.~\ref{FW_FW}, we build a CGH which is optically reconstructed using a SLM. The CGH is calculated by varying the coefficient of reflection of the SLM medium using the following amplitude function
\begin{equation}
H\left( x,y\right)= \frac{1}{2}\left\lbrace \beta\left( x,y\right) +a\left( x,y\right)cos\left[\phi\left( x,y\right)-2\pi \left(\xi x + \eta y \right)  \right] \right\rbrace \,,
\label{transmission}
\end{equation}
where $a\left( x,y\right)$ is the amplitude and $\phi\left( x,y\right)$ is the phase of the complex field, $\left(\xi,\eta \right)$ is the spatial frequency of the plane wave used as reference and $\beta\left( x,y\right)=\left[1+a^{2}\left(x,y \right)\right]/2 $ is the function bias taken as a soft envelope of the amplitude $a\left( x,y\right)$. The plane wave of reference is off--axis and introduces frequencies that separate the different orders of the encoded field.
\\

\begin{figure}[htbp]
 \centering
  \includegraphics[width=12cm]{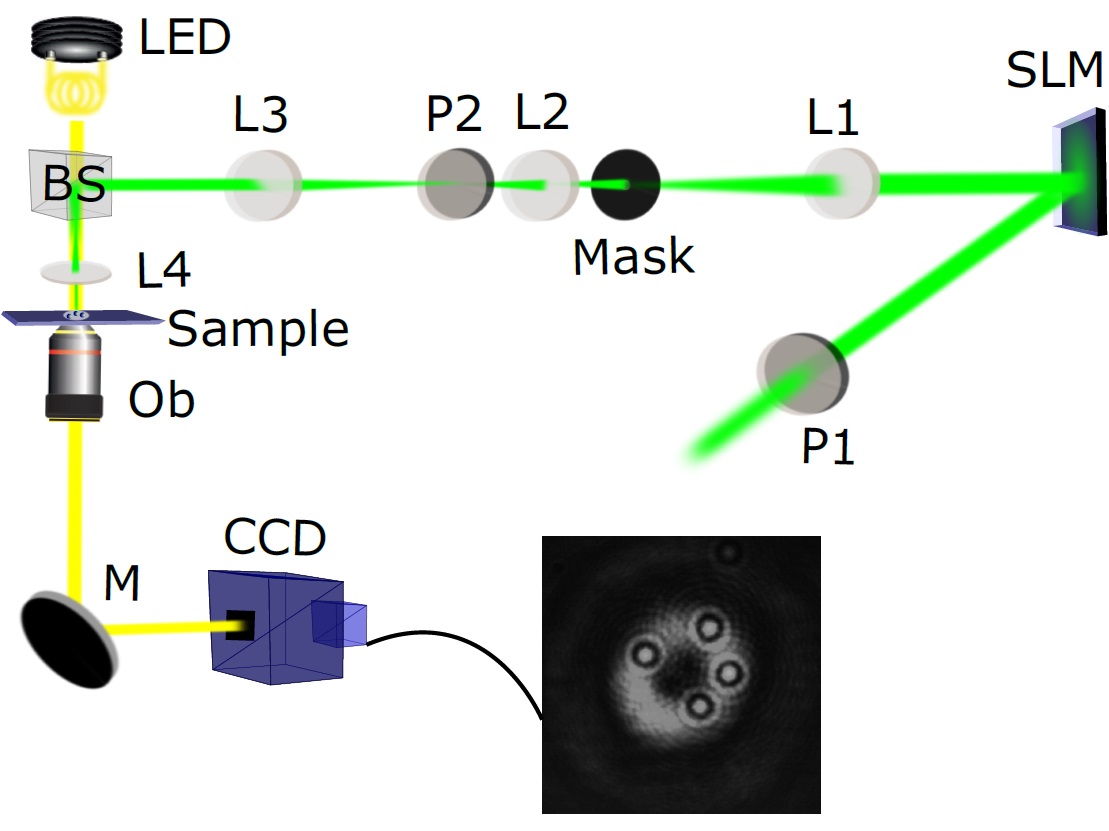}
 \caption{Schematic setup of the experimental holographic optical tweezers prototype for optical trapping using FWs. Using an Argon laser ($\lambda = 514.5nm$), with lenses (L), mirrors (M), beam splitter (BS), polarizers (P), a high numerical aperture objective (Ob, dichroic mirror (D), SLM is the reflection modulator and CCD is the camera for image acquisition ~\cite{suarezexperimental2020,suarez2021hotaab}.}
 \label{Setup_Pinza}
\end{figure}

\textbf{Holographic optical tweezers setup}. For the development of the holographic optical tweezers setup, shown in Fig.~\ref{Setup_Pinza}, was used an Argon laser with wavelength $\lambda = 514.5\,\mathrm{nm}$ and output power of $ 300\,\mathrm{mW} $. Initially, the beam passes through the spatial filter where it is expanded and 
is directed  to the SLM (LETO, Holoeye Photonics, with each pixel measuring $6.4~\mu\text{m}$ in a liquid crystal display o $1920\times1080$) which is placed at the input plane (Focus of lens L1)  .

In this system, to obtain an amplitude modulation of the FW beams, a computer--generated hologram was employed,  using Eq.~\ref{transmission}, and it is implemented in the SLM. The optical axis of the polarizer P1 is aligned at an angle of $ 0^{\circ} $ and of the polarizer P2 at $90^{\circ} $ in relation to the $y$--axis of SLM. The hologram is reconstructed due to the beam diffraction in the hologram CGH in the SLM. The reconstructed beam passes through two $4f$ systems to select the SLM diffracted beam and reduce the beam size, respectively~\cite{suarez2021hotaab}. 

The first $4f$ system consists of two lenses, L1 and L2 with focal lengths $150\mathrm{mm}$ and $50 \mathrm{mm}$, respectively. In the focal plane of the L1 lens, a mask was placed to select the different diffraction orders in the Fourier plane of the holographically reconstructed beam. The second $ 4f $ system was formed with two L3 and L4 lenses with focal lengths $150\,\mathrm{mm}$ and $25\,\mathrm{mm}$, respectively. A beam splitter (BS) reflects the beam vertically on the sample. The sample is placed in the focal plane of the L4 lens. To form the trapped particle image, the LED illumination system was used along with a $100\times$ and $ NA=1.25$ microscope objective.
\\

\h {\em\bf 4. Experimental Results}

\textbf{Propagation analysis of higher--order FW passing through ABCD paraxial optical system}. Initially, to study the effect of the $4f$ system on the propagation properties of the higher--order FW, we investigate the dynamical propagation for different values of $f_{2}$ where $f_{1}~=~150$~mm is fixed. For the optical generation of higher--order FW was used the experimental holographic setup in the Ref. ~\cite{Vieira2012,Vieira2015,Suarez2016}. 

As interesting intensity patterns, we consider that the longitudinal pattern of the FWs, generated in the $ z=z_{\text{SLM}}$ plane, is concentrated on a cylindrical surface. That is, we use a superposition of $(2N+1)$ Bessel beams with $\nu=2$ and whose longitudinal pattern in the range $0\leq z \leq L$ is given by a step function,  

\begin{equation}
F(z)= \left\{ \begin{array}{lc}
             1, &   \text{para} \quad  l_{1}\leq z \leq  l_{2} \\
             0, &   \text{elsewhere}\,.
             \end{array}
   \right.
\label{FW_Degrau_Exp}
\end{equation}
where $l_{1}=5~\text{cm}$ and $l_{2}=20~\text{cm}$.

We use the following values for the different parameters of the FWs in the input plane: $\lambda=632.8~\text{nm}$, $Q=0.9999898k$ and $L=50~\text{cm}$ thus obtaining a maximum number $ N=8$ , which implies in a of spot $\Delta \rho=53.51~\mu \text{m}$. Values of $f_{2}$ of $100~\text{mm}$, $150~\text{mm}$, and $200~\text{mm}$ were used to characterize the parameters of the beam passing through two convex lenses of different focal length points and validate the predicted theoretical results. 

\begin{figure}[htbp]
 \centering
  \includegraphics[width=14cm]{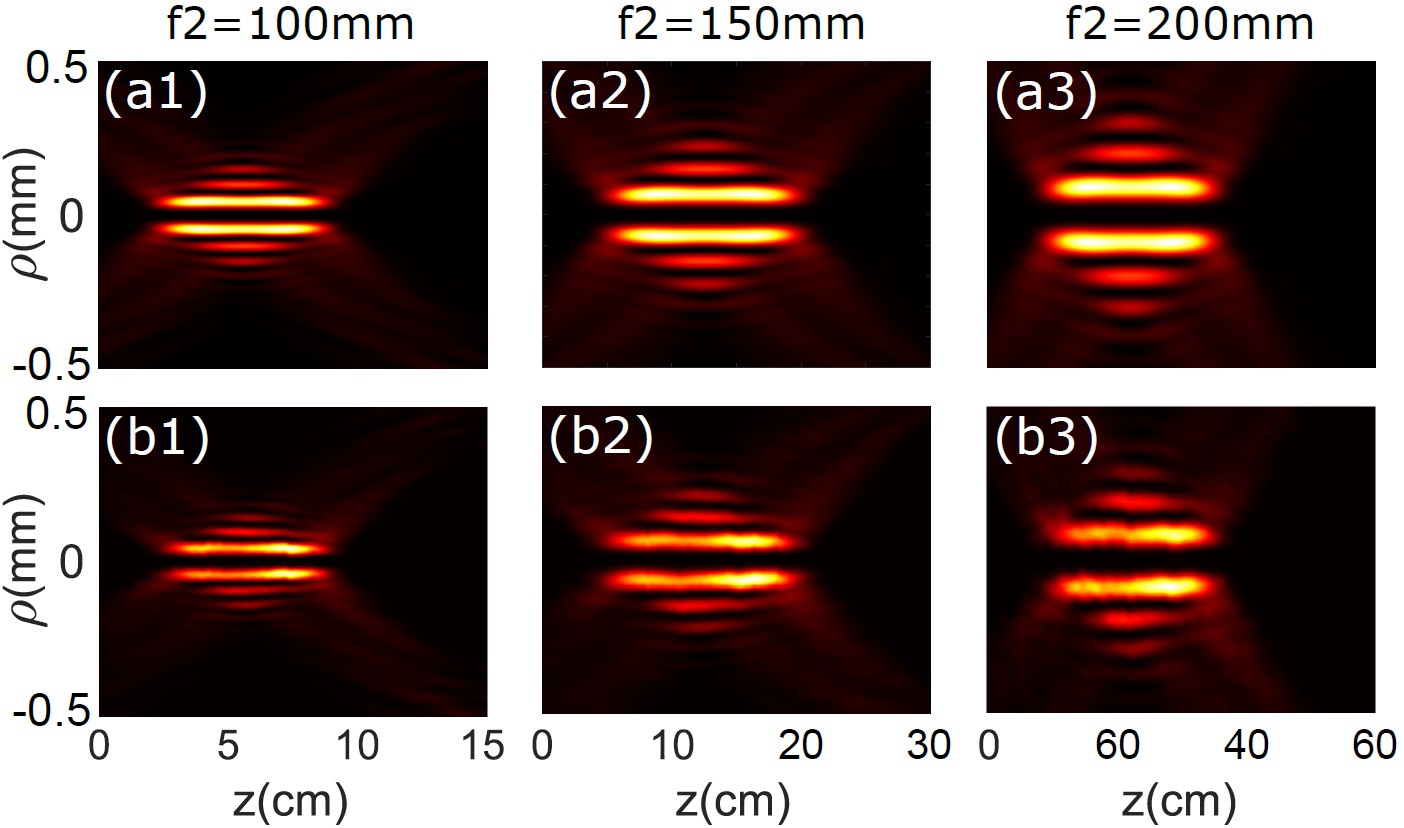}
 \caption{Dynamic propagation of the intensity distribution theoretical (up) and experimental (down) for FW of order 2 after passing through a $4f$ system for different focal lengths.}
 \label{Dinamic_propagation}
\end{figure}

\begin{figure}[t]
 \centering
  \includegraphics[width=14cm]{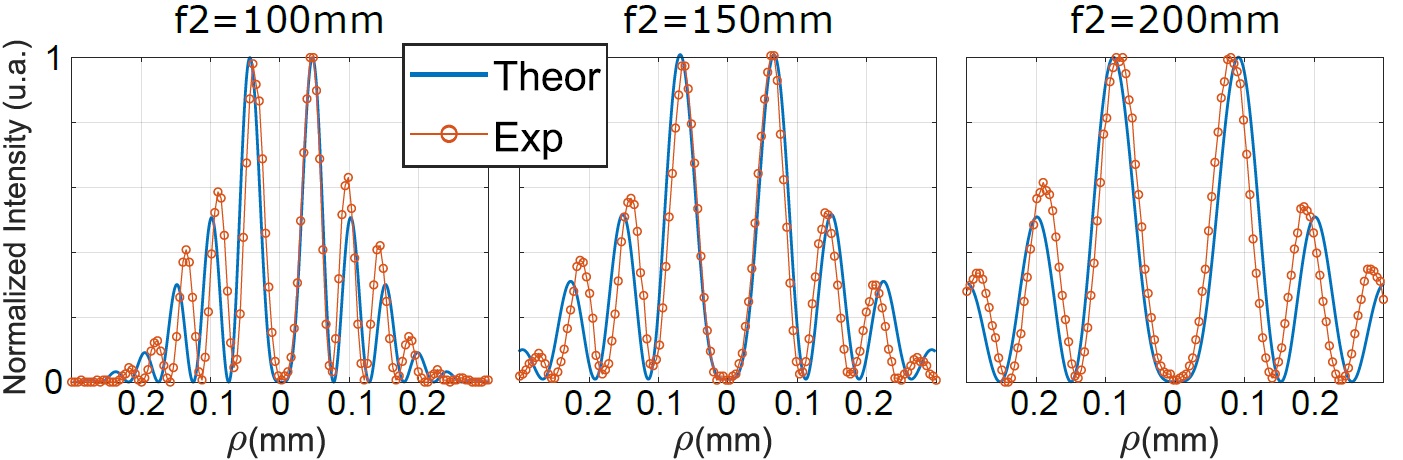}
 \caption{Comparison of the normalized intensity transverse profile between theoretical and experimental FW of order 2 when it passes through a system $4f$ for different focal lengths.}
 \label{Perfil_Transversal}
\end{figure}

The dynamic propagation of the theoretical intensity distribution  Fig.~\ref{Dinamic_propagation} $(a1)$--$(a3)$ obtained using the~\eqref{FW_OT_ABC}, and experimental Fig.~\ref{Dinamic_propagation} $(b1 )$--$(b3)$, at different focal lengths $f_{2}~=~(100,150,200~\text{mm})$ are shown. We can see that the FW is unchanged when it passes through the lens $f_{2}=150\text{mm}$, that is, the FW in the input plane (SLM placed in the focal plane of the lens L1) is exactly equal to the FW in the output plane (focal plane of lens L2). On the other hand, when the FW passes through the lens~$f_{2}=100~\text{mm}$ and $f_{2}=200~\text{mm}$ we have $\Delta~\rho~=~35.67~\mu\text{m}$ and $\Delta \rho=71.35~\mu \text{m}$, respectively. However, the longitudinal pattern will be set in the range $(2.2~\text{cm}~\leq~z~\leq~8.9~\text{cm})$ for $f_{2}=100~\text{mm}$ and $(8.9~\text{cm} ~\leq~z~\leq~35.5~\text{cm})$ for $f_{2}=200~\text{mm}$. In Fig.~\ref{Perfil_Transversal} we see the transversal profile of normalized intensity in the planes marked by the drawn lines, theoretical and experimental, for different focal lengths $f_{2}=(100,150,200~\text{mm })$. In this case, the reliability of the experimental result in relation to the theoretical prediction is evident~\cite{vittorino2019,suarezexperimental2020}.

\textbf{Optical trapping with of higher--order FW}. For optical trapping of micro--particles, we used the experimental setup of the Fig.~\ref{Setup_Pinza}. The sample consists of polystyrene micro--particles of radius $R=1.03~\mu \text{m}$ in water, between two thin microslide glass plates.

To reduce the dimensions of the beam, it passes through two systems $4f$. When the FWs optically reconstructed by SLM pass through this system, their propagation properties change according to the~\eqref{generalized_Huygens_Fresnel_diffraction_integral_FW2}. This revel that the spot radius is scaled by the factor  $(f_{2}f_{4}/f_{1}f_{3})$, where the longitudinal intensity pattern is scaled by $(f_{2}f_{4}/f_{1}f_{3})^2$.

Considering that the longitudinal pattern of the FWs, generated in the $ z=z_{\text{SLM}} $ plane, is given by for~\eqref{FW_Degrau_Exp} with $l_{1}=10~\text{cm}$ and $l_{2}=20~\text{cm}$. For the purpose respect the SLM resolution limit~\cite{Vieira2012,Vieira2015,suarezexperimental2020}, We use the following values for optical trapping in the input plane $z=z_{\text{SLM}}$: $\lambda=514~\text{nm}$, $Q=0.999993k$, $L=2~\text{m}$, thus getting a maximum number $N=27$, which implies a spot radius $\Delta \rho=52.58 \mu \text{m} $. 

\begin{figure}[htbp]
 \centering
  \includegraphics[width=10cm]{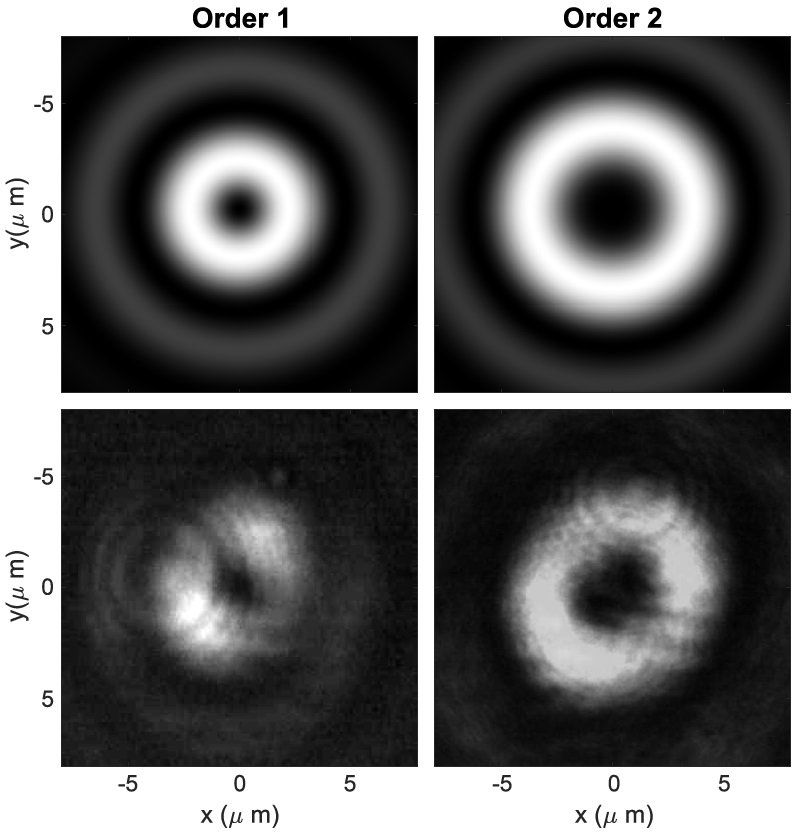}
 \caption{Cross section intensity profile for theoretical (top) and experimental (bottom) after passing through the two $4f$ systems for different order.}
 \label{Tansversal_Profile}
\end{figure}

\begin{figure}[t] 
 \centering
  \includegraphics[width=12cm]{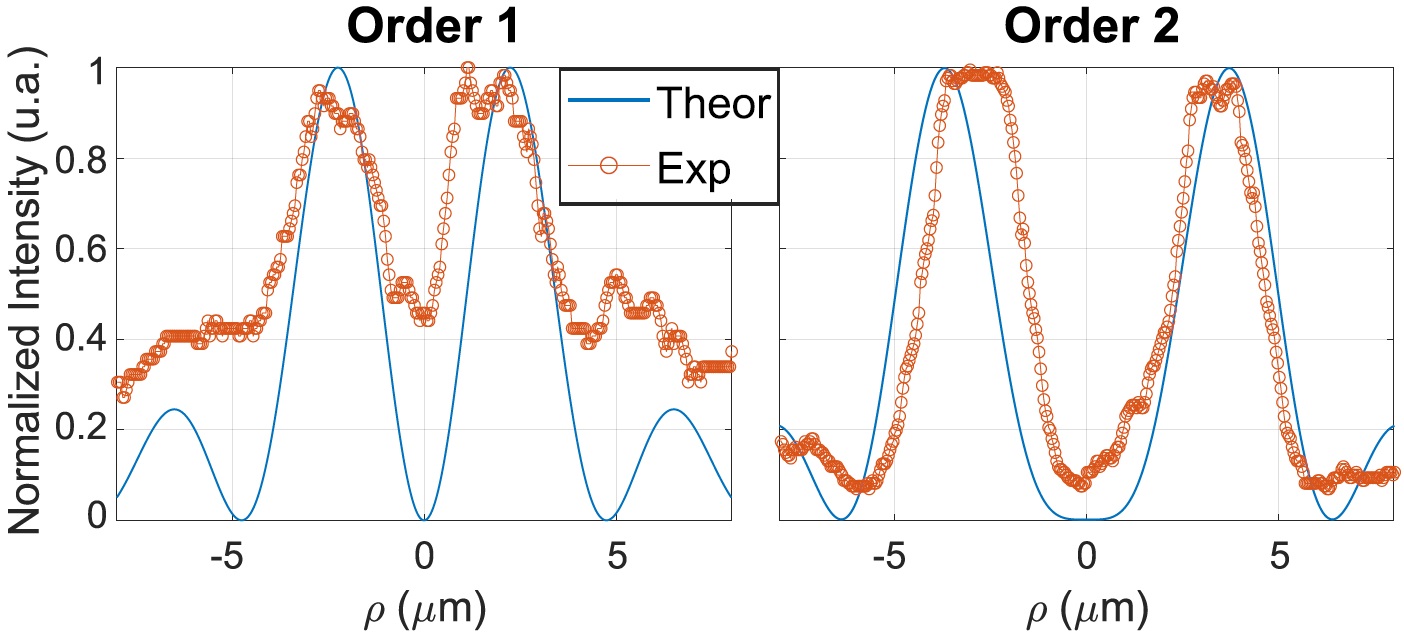}
 \caption{Comparison of the normalized cross section intensity over the plane $z_{c}$ between the theoretical and the experimental
result after passing through the two $4f$ systems for different order. $(a)$ $\nu=1$ and $(b)$ $\nu=2$.}
 \label{Transversal_Pattern}
\end{figure}

Fig.~\ref{Tansversal_Profile} shows the cross section intensity theoretical (top) and experimental (bottom) in the center of the step, plane $z_{c}=470~\mu \text{m}$, when the FWs passes through the $4f$ systems of the experimental arrangement in Fig.~\ref{Setup_Pinza} for different order $\nu=1$ and $\nu=2$. The normalized cross--section intensity over the plane $z_{c}$ are show in the Fig.~\ref{Transversal_Pattern}, where the size of spot radius is $\Delta_{\rho}=2.9~\pm~0.1~\mu\text{m}$. It is important to note that the size of the spot obtained experimentally is very close to the theoretical result.

Then, we were able to compare with the theoretical simulations of these beams and validate the reliability of the process of reducing the dimensions of the higher--order FW by the $4f$ systems in the holographic optical tweezers arrangement that we developed for the optical trapping.

\begin{figure}[htbp]
 \centering
  \includegraphics[width=14cm]{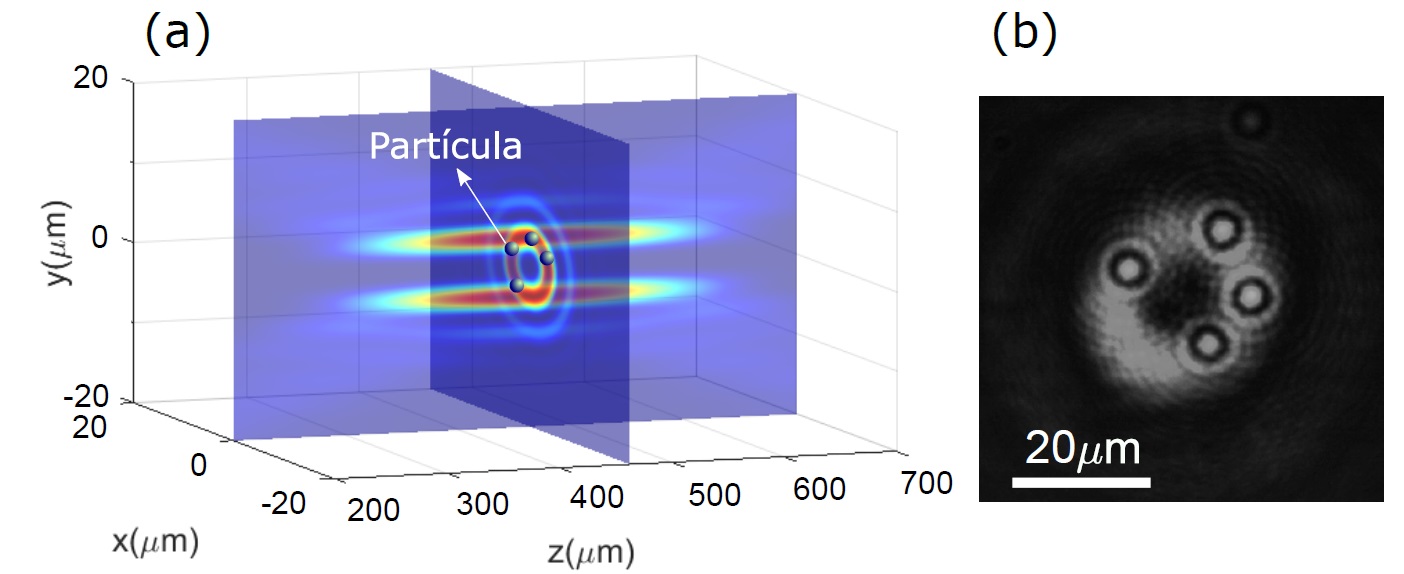}
 \caption{(a) Theoretical FW for $\nu=2$ generated after passing through the system $4f$ for trapping of micro--particles. Experimental optical trapping of multiple micro--particles in the plane $z_{c}$ for: (b) $\nu=1$ and (c) $\nu=2$. The power in the trapping plane was $1.2 \pm 0.5 $ \text{mW}}
 \label{FW_Trapping}
\end{figure}


Fig.~\ref{FW_Trapping} (a) shows the theoretical cross sections of the FW for $\nu=2$ generated for trapping after passing through the $4f$ system. The experimental results for micro--particles transversely trapped by the gradient force in the plane $z_{c}$, which corresponds to a stable equilibrium point, for $\nu=1$ and $\nu=2$ can be seen in Fig.~\ref{FW_Trapping} (b) and (c), respectively (Visualization 1). The power in the trapping plane was $2.5\pm 0.3$ \text{mW}. 

\begin{figure}[htbp]
 \centering
  \includegraphics[width=10cm]{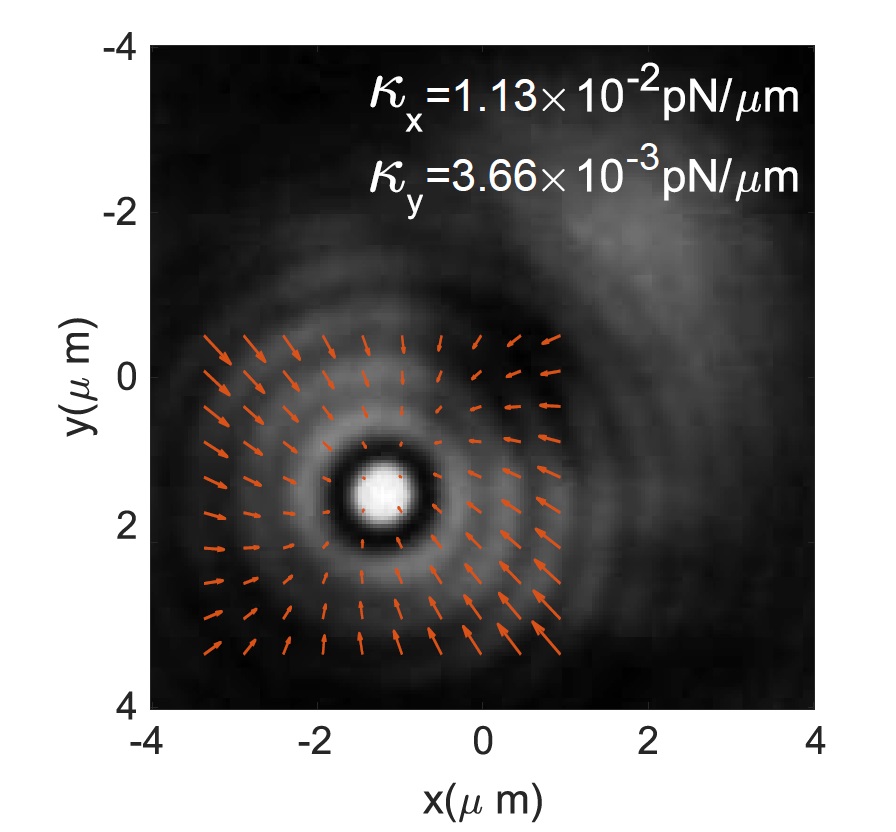}
 \caption{Force field reconstruction using the FORMA method, where $\kappa_{x}$ and $\kappa_{y}$ are the elastic constants along the $x$-– and~$y$-–axis, respectively for a FW of order 1.}
 \label{Froce}
\end{figure}

To determine the force distribution acting on the particle transversely trapping by the gradient force was used the FORMA method ~\cite{garcia2018high,suarezexperimental2020,suarez2021hotaab}. The particle movement around the equilibrium position and the force field distribution in the trapping plane is shown in Fig.~\ref{Froce} for a FW when $\nu=1$. The direction and magnitude of the arrows (orange) correspond to the direction and magnitude of the transverse force. The values of elastic (stiffness) constants $\kappa_{x}$ and $\kappa_{y}$ at the trapping plane are shown in the background of the figure.

To study how micro--particles are trapped and guided on transverse planes for different orders, we use a dynamic sequence of computer--generated holograms to create a dynamic scene~\cite{Vieira2015} and experimentally reproduce the movement of the beam at the transversal plane. For each values of $\nu$ we generate a hologram, in total 4 hologram for $\nu=0$ going up to $\nu=3$. The images are grouped and played back using the software (Slideshow player) of SLM--LETO. The CGH video was reproduced at rate of 10 frames/second.

\begin{figure}[htbp]
 \centering
  \includegraphics[width=14cm]{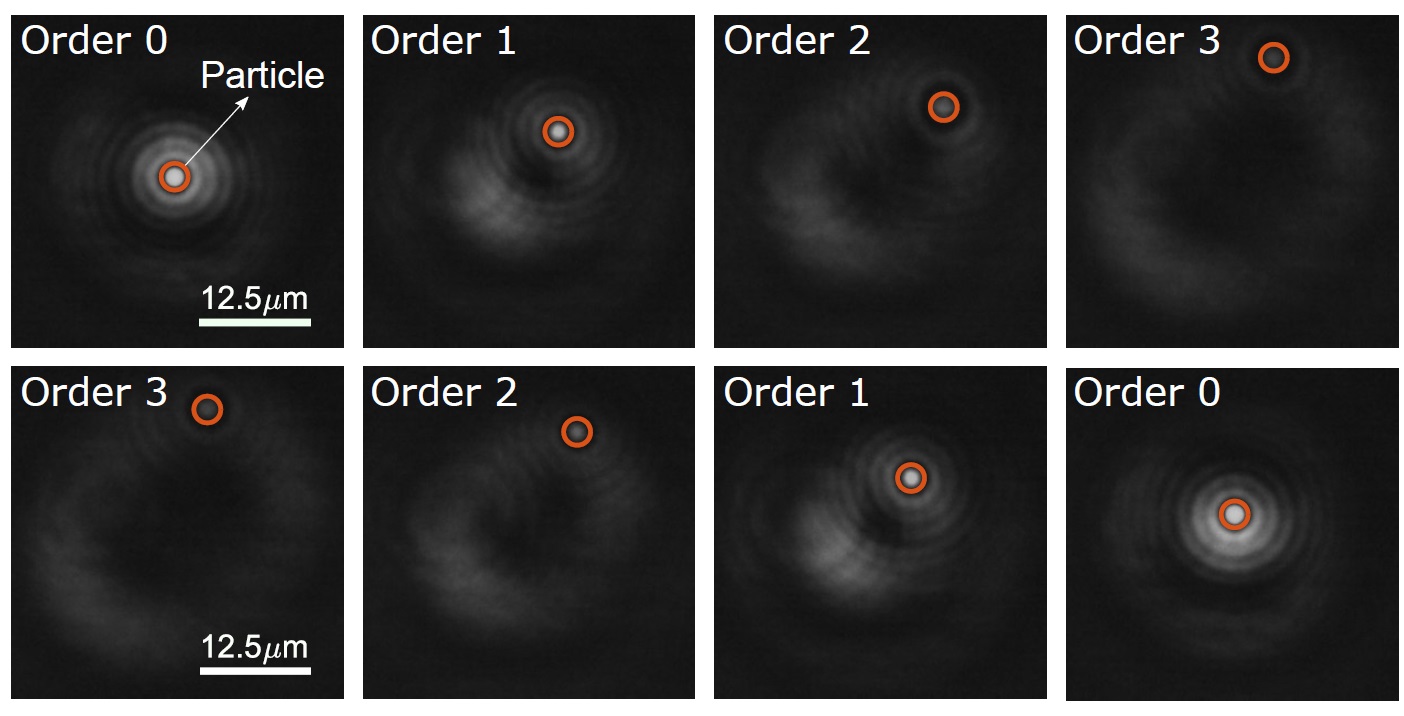}
 \caption{Trapping and guiding in the transverse plane of micro--particles (orange circle) using higher--order FWs ($\nu = 0,1,2,3)$, from computer--generated dynamic holograms. The power in the trapping plane was $1.2 \pm 0.5 $ \text{mW}.}
 \label{Dinamic}
\end{figure}

Fig.~\ref{Dinamic} shows the experimental result for trapping and guiding of micro--particles on the transversal plane using FWs of different orders. We can see how the particles is is moved away from the center following the maximum intensity as the FW order increases. On the other hand, the micro--particle returns to the center decreasing the FW order (Visualization 2). The perspective of this type of dynamic application is extremely promising and allows several possibilities for the use of optical trapping with structured light using holographic optical tweezers.  
\\

\h {\em\bf 5. Conclusions}  

This work presents the experimental optical trapping of micro--particles with higher--order Frozen Waves using holographic optical tweezers, for the first time to our knowledge. Stable trapping positions has been achieved, along with the characterization of its force distribution. Controlling different higher--orders FW, via dynamic holography, results in long, thin, and hollow cylindrical intensity profiles, which could control the guidance of micro--particles. The results obtained show that the optical trapping of micro--particles using high order FW presents excellent stability conditions. As well as promising and interesting possibilities interesting possibilities for dynamic trapping and guiding in a controllable way that can be applied in optical, biological, atmospheric sciences, among others.\\

\h {\em Acknowledgments.} The authors thank Leonardo A. Ambrosio and Michel Zamboni--Rached for their continuous collaboration. This work was supported by Federal University of ABC (UFABC); Coordenaç\~ao de Aperfeiçoamento de Pessoal de N\'ivel Superior-CAPES; Fundaç\~ao de Amparo a Pesquisa do Estado de S\~ao Paulo-FAPESP (16/19131-6); Conselho Nacional de Desenvolvimento Cient\'ifico e Tecnol\'ogico-CNPQ (302070/2017-6, 307726/2020-7).

\end{document}